\magnification=\magstep1
\hsize=13cm
\vsize=20cm
\overfullrule 0pt
\baselineskip=13pt plus1pt minus1pt
\lineskip=3.5pt plus1pt minus1pt
\lineskiplimit=3.5pt
\parskip=4pt plus1pt minus4pt

\def\negenspace{\kern-1.1em}



\newcount\secno
\secno=0
\newcount\susecno
\newcount\fmno\def\z{\global\advance\fmno by 1 \the\secno.
                       \the\susecno.\the\fmno}
\def\section#1{\global\advance\secno by 1
                \susecno=0 \fmno=0
                \centerline{\bf \the\secno. #1}\par}
\def\subsection#1{\medbreak\global\advance\susecno by 1
                  \fmno=0
       \noindent{\the\secno.\the\susecno. {\it #1}}\noindent}


\def\sqr#1#2{{\vcenter{\hrule height.#2pt\hbox{\vrule width.#2pt
height#1pt \kern#1pt \vrule width.#2pt}\hrule height.#2pt}}}


\newcount\refno
\refno=1
\def\y{\the\refno}
\def\myfoot#1{\footnote{$^{(\y)}$}{#1}
                 \advance\refno by 1}

\def\astfoot#1{\footnote{$^{(*)}$}{#1}}


\def\neq{\hbox{$\,$=\kern-6.5pt /$\,$}}





\newcount\secno
\secno=0
\newcount\fmno\def\z{\global\advance\fmno by 1 \the\secno.
                       \the\fmno}
\def\sectio#1{\medbreak\global\advance\secno by 1
                  \fmno=0
       \noindent{\the\secno. {\it #1}}\noindent}





\centerline{\bf{TIME EVOLUTION IN THE PRESENCE OF GRAVITY}} 
\vskip 1.0cm 
\centerline{by}
\vskip 1.0cm
\centerline{A. Pulido, A. Tiemblo\astfoot{To Miguel.} and R. Tresguerres}
\vskip 1.0cm
\centerline{\it {IMAFF, Consejo Superior de Investigaciones Cient\'ificas,}}
\centerline{\it {Serrano 113 bis, Madrid 28006, Spain}}
\vskip 1.5cm
\centerline{ABSTRACT}\bigskip 
We present a suggestion on the interpretation of canonical time evolution when gravitation is present, based on the nonlinear gauge approach to gravity. Essentially, our proposal consists of an internal-time concept, with the time variable taken from the dynamical fields characteristic of the nonlinear realization of the internal time-translational symmetry. Physical time evolution requires the latter symmetry to be broken. After disregarding other breaking mechanisms, we appeal to the Jordan-Brans-Dicke action, conveniently interpreted, to achieve that goal. We show that nontrivial time evolution follows, the special relativistic limit being recovered in the absence of gravity. 
\bigskip\bigskip 
\sectio{\bf Introduction}
\bigskip

Traditionally, dynamics dealt with the evolution laws of physical quantities in time. However, General Relativity (GR) conceives time itself (spacetime in fact) governed by dynamics. This vicious circle gives rise to difficulties in defining a satisfactory generalization of time evolution, valid also in the case when gravity is present. That is a central aspect --among others, certainly-- of what is called in the literature the {\it problem of time} in gravitational physics$^{(1)}$. On the other hand, most of the problems in quantizing gravity arise from the lack of a natural time variable at the classical level. As reviewed by Kuchar$^{(2)}$, the {\it multiple choice problem} and the related {\it Hilbert space problem} deal with the difficulty in deciding among the inequivalent quantum theories resulting from different choices of time. We claim that a revision of the classical concept of time is needed, in the context of canonical theories of gravitation, prior to quantization.

It is a common feature of generally covariant actions that they give rise to singular Hamiltonians. In particular, such Hamiltonians are linear in the scalar constraint, standing the latter as the generator of reparametrization invariance. Accordingly, in general, the Poisson bracket of any dynamical variable with a general-relativistic Hamiltonian gives rise to a symmetry transformation of the variable. Nevertheless, reparametrization invariance does not hinder physical time evolution. Actually, as far as a suitable time variable can be identified among the dynamical degrees of freedom of the theory, it remains possible to define time evolution as a field correlation, compatible with reparametrization invariance. Then, the consequence of the latter symmetry simply consists in that the time scale remains arbitrary. That can be easily seen in the case of the free particle$^{(3)}$. Being its Hamiltonian proportional to the mass shell condition $p^2-m^2\approx 0$, the time scale relating the affine time parameter and the proper time is measured by an arbitrary Lagrange multiplier. Thus, in fact, the lack of time evolution in GR, in its usual formulations, has a different origin. As we will discuss in the following, it derives from the absence of a suitable field to be identified as the time variable. Rovelli proposed to interpret dynamical evolution in terms of {\it evolving constants of motion}$^{(4)}$. In the present paper, we will not be concerned with time evolution in a quantized gravitational theory, but with the alternative attempt to develop a consistent {\it internal time} framework$^{(2)}$ at the classical level, as a necessary previous step towards quantum gravity.

According to the {\it internal time} point of view, a meaningful dynamical time evolution has to be evaluated with respect to a time variable taken from the field degrees of freedom of the theory itself. When internal time is identified with a given function, time evolution is evaluated as a correlation of the remaining dynamical fields with a subsystem of the full system, playing the role of a physical clock variable$^{(5)}$. In the search for a "good clock", it is usual to consider gravitation coupled to
matter$^{(2)(6)}$. The price to be paid is that matter clocks must be defined, for instance by proposing particular models of reference fluids, that are hardly conciliable with the idea of a model independent universal time. However, we will show that a natural internal time field can be identified from the gravitational variables themselves, when gravity is treated as the nonlinear gauge theory of a certain spacetime group. 

Our conception of time evolution in the presence of gravitation is summarized in the following three postulates. They constitute the guide for the present paper, where we will develop their consequences in terms of the nonlinear Poincar\'e gauge theory (PGT) of gravity. As our first postulate, we claim that time evolution exists as a physical fact. The necessity of this postulate will become clear in the following. Implicit in it, it is to be understood that whatever original {\it time symmetry} may be present in the action governing spacetime dynamics, it has to be broken in order to yield physical time evolution (since symmetry changes are not physically real). Further, according to the {\it internal time} concept, our second postulate enunciates that time evolution consists of a correlation between physical fields. Obviously, that requires a particular field to be chosen as the common clock reference; a role that cannot be played by the nondynamical affine parameter underlying reparametrization invariance (neither after symmetry breaking). In order to guarantee universality and model independence, the clock field --when possible-- should be a gravitational field rather than a matter field. Finally, the third postulate requires that, by switching off gravity, the special relativistic limit must be recovered. 

As we will see, the two latter postulates are satisfied by choosing the clock field to be identical with a dynamical time-like field provided by the nonlinear approach to PGT$^{(7)(8)}$. It is associated to internal time-translations possesing both, dynamical character, and features analogous to those of Minkowski coordinates. It is what we will introduce later as the time {\it Goldstone-coordinate}. This time variable reduces to the ordinary time coordinate of Special Relativity in the absence of gravitation (as required by the third postulate). In order that the first postulate also be satisfied, a breaking mechanism is needed, operating suitably to yield nontrivial time evolution. That will reveal to be highly nontrivial, so that the paper is mainly devoted to explain the related difficulties, and to propose a possible solution for them. 

The present work rests on several previous papers in
which we studied the gauge approach to gravity, and its
Hamiltonian formulation$^{(7)-(9)}$. As extensively discussed there, 
the key to construct a local gauge--invariant internal time is
provided by the nonlinear realization (NLR) of a given spacetime
group$^{(10)}$, in the context of the gauge approach to
gravity$^{(11)(12)}$. In particular, we deal with the nonlinear 
Poincar\'e gauge theory (NL-PGT) proposed by us in a previous
paper$^{(8)}$. The resulting formalism is expressed in terms of 
connection variables, closely related to the Ashtekar
ones$^{(13)}$. The coframes are identical with the nonlinear
translational connections. In our approach, the role of dynamical time will be related to the time component $\vartheta ^0$ of the coframe. (Its internal structure will be commented below.) We remark the relevance of the nonlinear gauge treatment of Gravitation as the most convenient tool to discuss the group origin of the dynamical time as 
the (nonlinear) time component of the translational connection, as discussed by us elsewhere$^{(7)}$.

The paper is organized as follows. In section 2 we present a short review of the nonlinear gauge realization of the Poincar\'e group, and in section 3 we discuss the role, as internal time, of the properly NLR-fields that we call the {\it Goldstone-coordinates}. Section 4 is devoted to show how the internal time-translational symmetry is actually present in ordinary gravity as a hidden symmetry. Next, in section 5, different attempts to break down the time symmetry are studied. We disregard all of them due to their unability to yield a satisfactory characterization of evolution with respect to internal time. Such characterization is presented in section 6, in terms of the Jordan-Brans-Dicke action. By identifying the scalar field in it with the internal time variable, a time-symmetry breaking mechanism results. The final remarks present a brief summary of the paper, and an additional discussion in the limit of vanishing gravity. 
\bigskip\bigskip 

\sectio{\bf Nonlinear coset realization of the Poincar\'e group}
\bigskip 
The present paper rests on the nonlinear realization (NLR) of 
the Poincar\'e group as proposed by us in previous works$^{(8)(9)}$. 
We refer to them and to the general literature on nonlinear
realizations$^{(10)}$ for further details. In the following, 
we restrict ourselves to give a short review of the main 
results, which are necessary to follow our discusion and to 
fix the notation.  

Let us consider a Lie group $G$ with a subgroup $H\,$. The right action of the subgroup $H$ on $G$ gives rise to a complete partition of the group manifold $G$ into equivalence classes, namely the left cosets $gH$. The elements of the quotient space $G/H\,$ are labeled by continuous coset parameters, say $\xi\,$. The nonlinear coset realization of $G$, with classification subgroup $H$, rests on the following definition of the left action of the transformation group $G$ on its own group manifold. Given $g\in G$ and $h\in H$, let $g$ act on the zero sections $\sigma\left(
\xi\,\right)$ as 
$$L_g\circ\sigma\left( \xi\,\right)= R_h\circ\sigma\left( \xi '\,\right)\,\,,\eqno(\z)$$
with $\xi '$ as a transformed coset parameter, (see (2.6) below, where an example of infinitesimal transformation $\xi ' =\xi +\delta\xi $ is calculated in a particular case). The only additional feature we need to know to our present purpose is that the nonlinear connection relates to the ordinary linear one $\Omega$ as
$$\Gamma =\,\sigma ^{-1}\left(d\,+\Omega\,\right)\sigma\,.\eqno(\z)$$
Only the components of $\Gamma$ involving the generators of $H$ 
behave as true connections, transforming inhomogeneously,
whereas the remaining components transform as tensors with
respect to the subgroup $H\,$. This is a main feature of
nonlinear realizations. 

We now apply this treatment to the foundation of gravitational gauge theories. The kind of gauge theories of spacetime groups we have in mind is that developed mainly by Hehl$^{(12)}$, which scarcely differs from the standard Yang-Mills approach. The principal reason for invoking NLRs is that they provide true tetrads transforming as covectors, without further {\it ad hoc} modifications of the linear translational connections, as it is the case in the linear approach$^{(14)}$. Let us consider in particular the nonlinear Poincar\'e gauge theory (PGT). Accordingly, we take the Poincar\'e group $G$ as the gauge group of gravitation, with $H$ chosen to be the Lorentz group. (Other choices of $H$ are possible$^{(8)}$, but we will not discuss this point here.) Being the Poincar\'e algebra given by the Lorentz generators $L_{\alpha\beta}$ and the translational 
generators $P_\alpha\,$ $(\alpha\,,\beta =\,0,...3\,)$, we identify the infinitesimal group elements $g$ in (2.1) to be 
$$g =\,e^{i\,\epsilon^\alpha P_\alpha }
e^{i\,\beta ^{\alpha\beta}L_{\alpha\beta}} 
\approx\,1+i\,\left( \epsilon^\alpha P_\alpha +\beta ^{\alpha\beta}L_{\alpha\beta},\right)\,.\eqno(\z)$$
In addition, the (infinitesimal) group elements of the right
acting Lorentz group $H$ are taken to be 
$$h =\,e^{i\,u^{\alpha\beta}L_{\alpha\beta}} 
\approx\,1+i\,u^{\alpha\beta}L_{\alpha\beta}\,,\eqno(\z)$$
and the sections are parametrized as  
$$\sigma =e^{-i\,\xi ^\alpha P_\alpha}\,,\eqno(\z)$$
being $\xi ^\alpha \,$ the (finite) coset parameters. We remark from now on the central role that the latter fields, arising from the nonlinear treatment of the translations, will play in our approach. Their meaning will be discussed in the next section. 

Substituting (2.3-2.5) into (2.1), an easy computation$^{(7)}$
yields the variation of the translational coset parameters 
$$\delta \xi ^\alpha =-\beta _\beta {}^\alpha\,\xi ^\beta 
-\epsilon ^\alpha\,,\eqno(\z)$$ 
(with $\beta _\beta {}^\alpha =u_\beta {}^\alpha$) showing that they transform exactly as Minkowskian coordinates. Let us now introduce the nonlinear gauge fields. The ordinary linear Poincar\'e connection $\Omega$ in (2.2), including translational and Lorentz contributions ${\buildrel (T)\over{\Gamma ^\alpha}}$ and $\Omega ^{\alpha\beta}$ respectively, reads 
$$\Omega :=-i\,{\buildrel (T)\over{\Gamma ^\alpha}} P_\alpha 
           -i\,\Omega ^{\alpha\beta}L_{\alpha\beta}
\,.\eqno(\z)$$
According to (2.2), we define the nonlinear connection, with values
on the whole Poincar\'e Lie algebra, as   
$$\Gamma :=\,\sigma ^{-1}\left(d\,+\Omega\,\right) \sigma 
          =-i\,\vartheta ^\alpha P_\alpha 
           -i\,\Gamma ^{\alpha\beta}L_{\alpha\beta}
\,,\eqno(\z)$$
being the nonlinear translational connection components 
$$\vartheta ^\alpha :=\,D\,\xi ^\alpha  
+{\buildrel (T)\over{\Gamma ^\alpha}}\,,\eqno(\z)$$ 
whereas the nonlinear Lorentz connection coincides with the linear one in this case; that is, $\Gamma ^{\alpha\beta} =\Omega ^{\alpha\beta}$. The Lorentz coframe (2.9) behaves as a Lorentz covector. (An alternative choice of $H=SO(3)$ allows to split the four--dimensional representation $\vartheta ^\alpha$ of the coframe into the SO(3) singlet $\vartheta ^0$ plus the SO(3) covector $\vartheta ^a\,$, but we will not discuss these details here. The interested reader is referred to Ref.(8).)  
\bigskip\bigskip

\sectio{\bf Translational coset parameters as Goldstone fields transforming like Cartesian coordinates. Internal time.}
\bigskip 

From the previous approach to the Poincar\'e group follows a result which holds for any spacetime group including translations, when taken as the dynamical group of the nonlinear gauge approach to gravity. For such groups, we claim that the tetrads $\vartheta ^\alpha$ are to be identified with the nonlinear translational connections, constructed from the linear translational connections ${\buildrel (T)\over{\Gamma ^\alpha}}$
and the translational coset fields $\xi ^\alpha $ (playing the role of Goldstone fields), as shown in (2.9). The variations under local translations read 
$$\delta {\buildrel (T)\over{\Gamma ^\alpha}}=\,D\epsilon ^\alpha 
\,,\qquad\delta\xi ^\alpha =-\epsilon ^\alpha\,,\eqno(\z)$$
compare with (2.6). Observe that, in the tetrad (2.9), the inhomogeneous translational variation of the linear connection ${\buildrel (T)\over{\Gamma ^\alpha}}$ is automatically compensated by that of $D\xi ^\alpha $. Actually, the structure (2.9) of $\vartheta ^\alpha $ may be interpreted as the Poincar\'e covariant differential of the coset field $\xi ^\alpha$, transforming on the one hand as a Lorentz tensor, and being, on the other hand, translationally invariant. Regarding translations, that situation corresponds rigorously to what is called in the literature the unitary gauge, where the Goldstone fields become rearranged into a redefinition of the corresponding gauge fields, being absorbed in the new dynamical variables by means of a redefinition isomorphic to a gauge transformation, with the group parameters replaced by the Goldstone fields $\xi ^\alpha $ themselves. (Notice, in fact, that the structure of the definition (2.2) is formally the same as that of a gauge transformation.)

In the Poincar\'e gauge theory (PGT) of gravitation, the Poincar\'e group plays the role of the internal symmetry. Accordingly, the translational coset fields $\xi ^\alpha $ (translations being a subgroup of the internal group) are dynamical fields of the theory. However, we point out the remarkable fact that, simultaneously, they transform as Cartesian coordinates, see (2.6). {\it Isomorphic} roughly meaning {\it equal}, one concludes that the $\xi ^\alpha $ are Cartesian coordinates. However, they are simultaneously translational Goldstone bosons, that is, dynamical objects. Then we have fields of the theory behaving as coordinates. Let us call them Cartesian {\it Goldstone-coordinates}. 

Obviously, being the {\it Goldstone-coordinates} dynamical fields,
they are completely different from the coordinates $x^i$ of the
underlying manifold. The latter are a non-dynamical tool of the
theory, whose function is that of implementing the idea of
locality. They are a mere label for characterizing the observers
(that is, the reference frames). On the contrary, as referred to
each observer, the Goldstone fields $\xi ^\alpha $ allow to
interpret spacetime measurements as values of a physical field.
This notion of observable space as something dynamical,
resembles the idea of a fluid of reference or {\it aether}, with
the essential difference that the dynamical field characterizing
each spacetime point is not a material {\it medium} introduced
by hand, but a field derived from a pure gravitational theory. 

Furthermore, the structure (2.9) of the tetrads clarifies, {\it \`a
la} gauge, the transition to the gravitation-free limit, namely
to special-relativistic flat Minkowski spacetime. Actually,
such transition results from the vanishing of the gravitational
connections, or equivalently from reducing the local symmetry to
the global one. (That is exactly analogous to switching off
electromagnetism by putting the electromagnetic potential to
zero.) Since the tetrads do not coincide with standard linear
connections, the vanishing of the latter is compatible with the
non-vanishing of the tetrads, which, in the transition to the
non-local realization of the Poincar\'e group, reduce to  
$$\vartheta ^\alpha = d\xi ^\alpha\,.\eqno(\z)$$
We recognize the usual form of tetrads in Special Relativity,
expressed in terms of the Cartesian {\it Goldstone-coordinates}.
Notice that, in the global case, the variations of $\xi ^\alpha
$ are Poincar\'e ones with constant group parameters, so that no
connections are necessary to cancel out inhomogeneous
contributions. In fact, in the absence of gravity, the Cartesian
{\it Goldstone-coordinates} become indistinguishable from the
standard Cartesian coordinates. The spacetime of Special Relativity 
thus reveals to have a dynamical origin. It constitutes the remaining 
structure after switching off the connections in a nonlinear gauge 
theory of gravity. 
\bigskip\bigskip 

\sectio{\bf Hidden translational symmetry in ordinary gravity.}
\bigskip 

Before entering the possible application of the previous considerations to gravitational actions, let us make some general remarks on the meaning of dynamical time evolution. First of all, if one accepts the course of time as an objective reality (remember that this is not an universal 
belief; according to several authors, time does not exist at
the fundamental level$^{(6)(15)}$), then one must clearly distinguish 
it from a symmetry transformation. Actually, symmetry transformations 
relate physically equivalent descriptions to each other. Thus, if 
time evolution were merely a transformation under a certain {\it time 
symmetry}, it would reflect no real changes. Thus, the only
way to save time as an actual alteration is to break down the 
corresponding (local) symmetry. The breaking will provide physically distinct time values, and thus the necessary reference for real time running. We remark as our first postulate about time evolution the necessity of breaking {\it time symmetry} --whatever it may be--. 

Retaining this observation in mind, our first task is to identify 
what one should understand in the following as {\it time symmetry}. 
We will examine two main symmetries related in the literature to
the idea of dynamical evolution, namely reparametrization
invariance and time-translational invariance
respectively$^{(7)}$. They are conceptually quite different. The
former constitutes a non internal symmetry, concerning
transformations of a coordinate-like affine parameter (an
unphysical quantity, not included among the dynamical degrees of
freedom of the theory). Reparametrization invariance guarantees
the indistinguishability of physical descriptions with respect
to the rescaling of the fields on such affine parameter. On the
other hand, in the case of PGT and similar theories of gravity {\it \`a la Hehl}$^{(12)}$, translations are included in the gauge group as a constitutive part of the internal symmetry. In particular, we will focus our attention on time-translations. 

Certainly, as far as they remain unbroken symmetries, neither
reparametrizations nor time-translations can represent time
evolution. Nevertheless, let us make a comparison between both,
according to which, one of them is preferable as the {\it time
symmetry} to be broken in order to yield physical time. At this 
point, we invoke our second postulate about time as the criterion to guide the pertinent choice, namely: The course of time is to 
be understood as a correlation between physical fields. Accordingly, 
we examine both candidate symmetries. Regarding reparametrization 
invariance, we find that the only recognizable time-like parameter 
is an external, unphysical, affine one. Instead, in internal 
time-translations, a dynamical time-like field is involved, namely 
what we called in the previous section the time {\it Goldstone-coordinate}. 
Obviously, we expect the breaking of the latter, rather than that of 
the former, to yield the sort of field correlations to be identified 
as dynamical time evolution. Indeed, in this case, a dynamical field, 
and not the undynamical affine parameter, would constitute the 
necessary clock reference. For this reason, we propose to characterize 
time evolution as emerging from breaking the time-translational 
symmetry. 

Both already enunciated postulates do not suffice to identify the time variable in an unambiguous way. Actually, they could be satisfied for instance by a spatial {\it Goldstone-coordinate}, which is also able to define field correlations. Thus, in the search for a suitable time variable, a further specification is required. We propose our very simple and natural third postulate, according to which, in the absence of gravitation the special relativistic limit must be recovered. In view of the tetrad structure (3.2) resulting from switching off gravity, it is clear that the postulate favours the choice of the time {\it Goldstone-coordinate} $\xi ^0$ as the time variable. Actually, the breaking of the local time symmetry by putting the translational connection equal to zero leads to the special-relativistic tetrad component $\vartheta ^0 =d\xi ^0$, allowing to define dynamical time evolution in the context of the remaining forces. Such time variable, as a constitutive part of the tetrad, couples universally to any other field (to Dirac matter fields inasmuch as to gauge fields). Notice that, being time a gravitational variable, time evolution is to be regarded as a gravitational effect.

Recall that, in the usual Hamiltonian treatment of
gravity, reparametrization invariance is generated by the
scalar constraint. Besides it, let us pay attention to
the conjugate momentum $p$ of the time {\it Goldstone-coordinate}
$\xi ^0$, see (4.23) below. This {\it time momentum} will become
relevant to our purposes after breaking translational
time-invariance. Indeed, if our definition of physical time
evolution were satisfactory, in addition we should identify an
associated time operator (the analogue of the usual Hamiltonian),
whose action on physical observables would yield dynamical
evolution, in analogy to classical dynamics. One can anticipate that the time momentum $p$ could play this role. We will return back to this 
point later. 

Let us now study, in the framework of gravitational dynamics in
vacuum, the different aspects of reparametrization invariance
and time-translational invariance, and the consequences of
breaking the latter, which in principle one expects to give rise to real time evolution, as discussed above. We begin summarizing the ordinary
treatment of gravitational theory. For the sake of simplicity,
we are going to consider as ordinary gravity the
Samuel-Jacobson-Smolin Euclidean action$^{(17)}$, written in the form
$$S=\int d^4x\,\eta ^{abcd}\left[ v_a e_{bi} F_{cdi}
-{1\over2}\epsilon _{ijk} e_{ai} e_{bj} F_{cdk}\right]\,,\eqno(\z)$$
where $F_{abi}$ is the $SO(3)$ field strenght tensor, and
$(v_a\,,e_{bi})$ are the components of the usual $SO(4)$ tetrad.
The action (4.1) is invariant both with respect to the explicit
$SO(3)$ transformations 
$$\eqalign{\delta (M) A_{ai} &=-\left( \partial _a M_i +\epsilon
_{ijk} A_{aj} M_k\,\right)\cr 
\delta (M) v_a &=0\cr 
\delta (M) e_{ai} &=-\epsilon _{ijk} e_{aj} M_k\,,\cr }\eqno(\z)$$
and under the remaining transformations of $SO(4)\sim
SO(3)\times SO(3)$, namely   
$$\eqalign{\delta (L) A_{ai} &=0\cr 
\delta (L) v_a &=L_i e_{ai}\cr 
\delta (L) e_{ai} &=-L_i v_a +\epsilon _{ijk} e_{aj} L_k\,.\cr }\eqno(\z)$$
These symmetries allow to remove irrelevant degrees of freedom,
as we will see below. 

Next we perform a foliation of the spacetime manifold, yielding
a $3+1$ decomposition, so that equal-time spatial hypersurfaces
become defined by a constant value of a time-like affine
parameter, say $t$. In order to do so, one introduces a
congruence of curves with tangent $t^a$ (such that $t^a\partial
_a t =1$). Then, the Lie derivative along $t^a$ will represent a
sort of "time derivative" (denoted in the following by means of
a dot) with respect to the affine parameter $t$. With this
assumption, the resulting $3+1$ decomposition transforms the
action (4.1) into  
$$\eqalign{ S=\int dt\,\int d^3x\,\Bigl\{ &\eta ^{abc}\dot{A} _{ai}
\bigl[ 2v_b e_{ci} -\epsilon _{ijk}e_{bj}e_{ck}\bigr] 
+a_i D_a\bigl[ \eta ^{abc}\bigl( 2v_b e_{ci} 
-\epsilon _{ijk}e_{bj}e_{ck}\bigr)\bigr]\cr  
&+\hat{v} _{_0} \eta ^{abc} e_{ai} F_{bci}
-\hat{u} _i\,\eta ^{abc}\bigl[ v_a F_{bci} 
+\epsilon _{ijk} e_{aj} F_{bck}\bigr]\Bigr\}\,,\cr }\eqno(\z)$$
where we defined $a_i:=A_{0i}$, $\hat{u} _i := e_{0i}$, and for
later convenience we denoted by $\hat{v} _{_0}$ the time
component of $v_a$. The resulting expression (4.4) simplifies
drastically when rewritten in terms of suitable variables. To this purpose, we first introduce the inverse of the triad,
namely $e_i{}^a$ such that $e_i{}^a e_{aj}=\delta _{ij}$, and
accordingly we define $v_i:=e_i{}^a v_a$. Making then use of the
relation 
$$\eta ^{abc}:= e\, e_i{}^a e_j{}^b e_k{}^c \epsilon _{ijk}\,,\eqno(\z)$$ 
and of the matrix 
$$M_{ij}:=\delta _{ij}-\epsilon _{ijk} v_k\,,\eqno(\z)$$ 
the first contributions to (4.4) become expressible in terms of 
$$\eta ^{abc}\bigl( 2v_b e_{ci} -\epsilon
_{ijk}e_{bj}e_{ck}\bigr) =-2\,e M_{ij} e_j{}^a =:-E_i{}^a\,.\eqno(\z)$$
Here a redefined {\it vierbein} $E_i{}^a$ is introduced, which
relates to the original one as 
$$e_i{}^a ={1\over{2e}}\bigl( M^{-1}\bigr) _{ij} E_j{}^a\,,\quad 
e_{ai}= 2e E_{aj}M_{ji}\,.\eqno(\z)$$
The corresponding determinants $e:={\rm det} e_{ia}$ and
$E:={\rm det} E_{ia}$ relate to each other as
$\sqrt{2E(1+v^2)}=1/(2e)$. Taking all these definitions into
account, after a little algebra, the action (4.4) reduces to the expression 
$$S=\int dt\,\int d^3x\,\Bigl\{ -\dot{A} _{ai} E_i{}^a  
-a_i D_a E_i{}^a  + u^a C_a + v_{_0} S_{_0}\Bigr\}\,,\eqno(\z)$$
where the field $v_a$ is absent, as it corresponds to irrelevant
degrees of freedom eliminated by exploiting the symmetries of
the action, as pointed out before. In (4.9), we made use of the notation 
$$\eqalign { u^a :=&\,{{\hat{v}_{_0}v_i +
M_{ij}\hat{u}_j}\over{2e(1+v^2)}} E_i{}^a\,,\cr 
v_{_0} :=&\,{{\hat{v}_{_0}-v_i\hat{u}_i}\over{4e(1+v^2)}}
\,.\cr }\eqno(\z)$$
Although (4.9) suffices for discussing the particular points we are
interested in, let us be more rigorous going a step further, in
order to read out the information about constraints from the
usual Hamiltonian formalism. Starting from the Lagrangian ${\cal
L}$ implicit in (4.9) as $S=\int dt\, {\cal L}$, we define
the momenta  
$$\pi _i{}^a :={{\delta {\cal L}}\over{\delta \dot{A}_{ai}}}\,,\quad 
\pi _i :={{\delta {\cal L}}\over{\delta \dot{a}_i}}\,,\quad
\sigma _{ia} :={{\delta {\cal L}}\over{\delta \dot{E}_i{}^a}}\,,\quad
\sigma _a :={{\delta {\cal L}}\over{\delta \dot{u}^a}}\,,\quad
\Delta :={{\delta {\cal L}}\over{\delta \dot{v}_{_0}}}=0\,.\eqno(\z)$$ 
From all of them, the only nonvanishing one results to be  
$$\pi _i{}^a =-E_i{}^a\,.\eqno(\z)$$ 
Finally, we get the singular Hamiltonian  
$$H=\int d^3x\,\Bigl\{ a_i D_a E_i{}^a  - u^a C_a - v_{_0}
S_{_0} +\lambda _{ai}\bigl( \pi _i{}^a +E_i{}^a\bigr) +\lambda
_i \pi _i  +\mu _i{}^a \sigma _{ia} +\mu ^a \sigma _a +\xi
\Delta \Bigr\}\,.\eqno(\z)$$ 
As announced above, in the following, we will identify the
Hamiltonian (4.13) derived from the original Samuel-Jacobson-Smolin
action (4.1) as the standard scheme of gravitational theory. All
future modifications of the usual treatment will refer to the latter action as the necessary comparison term. In (4.13) we recognize the Gauss
constraint, which is a first class constraint given by the
$SO(3)$ covariant derivative 
$$D_a E_i{}^a :=\partial _a E_i{}^a +\epsilon _{ijk} A_{aj}
E_k{}^a \,,\eqno(\z)$$ 
being identified as the generator of $SO(3)$ rotations. Further,
the vector constraint reads 
$$C_a := E_i{}^b F_{abi}\,,\eqno(\z)$$ 
and the scalar constraint is defined as 
$$S_{_0}:=\epsilon _{ijk} E_i{}^a E_i{}^b F_{abk}\,.\eqno(\z)$$ 
The scalar constraint is the well known one playing the role of
generator of the reparametrization symmetry, that is of the
invariance of the action under rescalings of the time-like affine parameter $t$. 

In addition to the just mentioned symmetries, whose generators
are present in the action, from our previous discussion on
gravity as the nonlinear gauge theory {\it \`a la Hehl} of a
certain spacetime group$^{(12)}$, we know that a further
(internal) translational symmetry exists in the theory, even if
it is not manifest in the action (4.1) of ordinary gravity. The
reason for this symmetry not to show up in the standard approach
is that usually one assumes the tetrads to lack further internal
structure. In order to make explicit the hidden symmetry, in
particular the time-translational one, recall that according to
the NLR-approach, the time component of the tetrad, namely the
dynamical object displayed as $v_a$ in the action (4.1), is not a
simple field, but a more complex structure involving {\it Goldstone-coordinates} as discussed in previous section; in particular, it
involves the time {\it Goldstone-coordinate}, say $\xi ^0$, and the
time-translational connection, say $\Gamma _a$ (redefined here
with respect to that of (2.9) by including in it the connection part of the covariant derivative of $\xi ^0$, in order to simplify calculations and reasoning). Thus, we propose to write $v_a$ in the form 
$$v_a :=\partial _a \xi ^0 +\Gamma _a\,.\eqno(\z)$$ 
The addition of the {\it extra} degree of freedom $\xi ^0$ does not
modify the number of degrees of freedom of the theory, since the
translational symmetry is present, yielding the variations 
$$\delta\Gamma _a =\partial _a\epsilon ^0\,,
\qquad\delta\xi ^0 =-\epsilon ^0\,,\eqno(\z)$$ 
compare with (3.1), so that the total number of degrees of freedom 
is the same as in standard gravity. 

Only by making explicit the structure of $v_a$, the
time-translational symmetry becomes explicitly displayed. That
is what we are going to show in the following, showing that the theory remains ordinary gravity. Taking thus (4.17) into account, we
procede as before, foliating the action (4.1) so that it becomes  
$$\eqalign{ S=\int dt\,\int d^3x\,\Bigl\{ &-2\dot{A} _{ai}\,e M_{ij} e_j{}^a
-2 a_i\,D_a\bigl(e M_{ij} e_j{}^a\bigr)\cr 
&+\bigl(\dot{\xi ^0}+\hat{\Gamma} _{_0}\bigr)\,\eta ^{abc} e_{ai} F_{bci}
-\hat{u} _i\,\eta ^{abc}\bigl[ v_a F_{bci} +\epsilon _{ijk}
e_{aj} F_{bck}\bigr]\Bigr\}\,.\cr }\eqno(\z)$$ 
Replacing now, for convenience, the previous definitions (4.10) by 
$$\eqalign { u^a :=&\,{{\hat{\Gamma}_{_0}v_i 
+ M_{ij}\hat{u}_j}\over{2e(1+v^2)}} E_i{}^a\,,\cr 
v_{_0} :=&\,{{\hat{\Gamma}_{_0}-v_i\hat{u}_i}\over{4e(1+v^2)}}
\,,\cr }\eqno(\z)$$
the action (4.19) takes the form 
$$S=\int dt\,\int d^3x\,\Bigl\{ -\dot{A} _{ai} E_i{}^a  
-a_i D_a E_i{}^a  + u^a C_a + v_{_0}S_{_0}
+2\dot{\xi ^0}eE\bigl( S_{_0}+2\,\tilde{v}^a
C_a\bigr)\Bigr\}\eqno(\z)$$ 
instead of (4.9), compare with it. Here we introduced the notation 
$$\tilde{v}^a :=v_i E_i{}^a =2e\,e_i{}^a e_i{}^b\, v_b 
={1\over{2e}} E_i{}^a E_i{}^b v_b \,.\eqno(\z)$$
If we want to construct the Hamiltonian version of the theory,
we have to introduce, in addition to the momenta (4.11), a further
"time momentum" related to the time field $\xi ^0$, namely 
$$p :={{\delta {\cal L}}\over{\delta \dot{\xi ^0}}}\,.\eqno(\z)$$  
For what we are going to say, it is not necessary to write down
the Hamiltonian explicitly. It suffices to mention that, in the
Hamiltonian version, the main first class constraints turn out to be 
$$\eqalign {&D_a E_i{}^a \approx 0\cr  
&p-2eE\bigl( S_{_0}+2\,\tilde{v}^a C_a\bigr)\approx 0\cr 
&C_a \approx 0\cr 
&S_{_0} \approx 0\,. \cr }\eqno(\z)$$ 
Among them, a Schr\"odinger-like constraint is present, standing
besides the scalar constraint $S_{_0}$ (that is, besides the
generator of reparametrization invariance) as the generator of
the time-translational symmetry transformations. Due to the fact
that, on the other hand, $C_a$ and $S_{_0}$ still remain
separately the vector and the scalar constraint, respectively, being weakly equal to zero, the time momentum $p$ also vanishes. Thus, there is nothing new with respect to the standard theory studied above. The situation reduces to that of ordinary gravity. The only new element is the manifestation of
the time-translational constraint. Since all fields commute with
it, all of them are invariant under the corresponding symmetry. 

Let us summarize the result of the present section saying that,
provided one identifies the tetrads with the (nonlinear)
translational connections, ordinary gravity is to be seen as a
theory with translational invariance. This symmetry is hidden
due to the fact that the transforming fields are systematically
replaced by invariant combinations of them, which is
characteristic for the unitary gauge.
\bigskip\bigskip

\sectio{\bf Searching for nontrivial time evolution.}\bigskip 

As discussed in section 4, we disregard reparametrizations of the (unphysical) affine parameter $t$ as having anything to do with physical time evolution. In fact, our description of the physical course of time rests on the second kind of time-like symmetry considered by us, namely the (internal) time-translations. Transformations of this type represent changes of the time {\it Goldstone-coordinate} $\xi ^0$, no longer an external parameter like $t$, but a dynamical field of the theory. Recall that, for this reason, we pointed out such transformations as good candidates for picturing time evolution as a correlation between physical fields, as referred to $\xi ^0$. However, since no real time changes can be defined as far as the time symmetry is unbroken, the emergence of physical time evolution requires to break the symmetry. As long as time-translational symmetry remains unbroken, the time momentum $p$ stands as the generator of the corresponding symmetry transformations, its vanishing $p\approx 0$ constituting a first class constraint. On the other hand, the breaking will in general hinder the nonvanishing of $p$. Nevertheless, that does not automatically transform $p\neq 0$ into a physical time operator giving rise to nontrivial time evolution, due to the interplay between $p$ and the scalar constraint of reparametrization invariance, as we will see below. 

In view of our interpretation of time evolution as related to the breaking of time-translational symmetry, the usual statement on the
inexistence of time in GR at the fundamental level may be
reinterpreted as follows. Certainly, no real time evolution is
present in standard gravity. The reason for it is that
time-translations underly that theory as an unbroken symmetry.
Due to the existence of this (hidden) symmetry, the search for
real time evolution in ordinary gravity makes no sense.
Time-translational invariance, as well as reparametrization
invariance, are symmetries of GR, none of them describing real
time changes. 

Thus, the only way to get time evolution is to go beyond GR,
suitably modifying the ordinary theory. As we have seen, the 
explicit display of the translational invariance in ordinary 
gravity gives rise to the constraint (4.24b). Nevertheless, the 
independent vanishing of the time momentum $p$ and of the remaining contributions to the time-translational constraint separately, invalidates the possibility of interpreting (4.24b) as a Schr\"odinger-like scalar constraint associated to nontrivial evolution. We retain yet the result that an equation of the Schr\"odinger type accompanies the translational symmetry, and we propose to construct a time-translationally broken theory. By avoiding the vanishing of the time momentum $p$, we hope to get it to play a role similar to the Hamiltonian of standard classical mechanics, 
yielding physical time evolution with respect to the time 
{\it Goldstone-coordinate} as something different from a symmetry 
transformation. 

Let us study how to break the translational symmetry of the
ordinary theory in order to get nontrivial dynamical time
evolution. The most radical way to do it is to take the
translational connection $\Gamma _a $ to be zero in (4.17). A
similar, maybe somewhat more general result follows from
imposing the Frobenius foliation condition on the time component
of the tetrad. In the differential-form language of Refs.(8)(9), the Frobenius equation for the (invariant) component $\vartheta ^0$
may be written as $\vartheta ^0\wedge d\vartheta ^0 =0$. Its
general solution reads $\vartheta ^0 =u\,d\tau$. In the
component-language we are using in this paragraph, this result
means that $v_a =u\partial _a \tau$. The question now is how to
interpret this formal expression for $v_a$. In particular, is
$\tau$ to be taken as a nondynamical affine parameter, or as a
dynamical field instead? The former from both alternatives is
the usual one, leading to ordinary gravity$^{(8)}$. The reason is that the translational symmetry is respected in this case. Now we are interested in analyzing the latter possibility. Consequently, we take the solution of the Frobenius condition to be 
$$v_a =f(\xi ^0)\partial _a\xi ^0\,,\eqno(\z)$$ 
with $\xi ^0$ a dynamical field, identified in the following
with the already known time {\it Goldstone-coordinate} introduced by us in sections 2 and 3. Contrary to the previous case, the translational symmetry now becomes broken. Indeed, being $\xi ^0$ not invariant under time-translational transformations, see (4.18), the lack of a compensating translational connection, as in (4.17), avoids the invariance and even the covariance of $v_a$, as given by (5.1). Thus, replacing (5.1) in the original Samuel-Jacobson-Smolin action (4.1), it becomes 
$$S=\int d^4x\,\eta ^{abcd}\left[ f(\xi ^0)\partial _a\xi ^0
e_{bi} F_{cdi}-{1\over2}\epsilon _{ijk} e_{ai} e_{bj}
F_{cdk}\right]\,,\eqno(\z)$$
essentially differing from the original one, although formally analogous, due to the break down of the translational symmetry in (5.2). Now we
follow the same steps as in the ordinary theory. We find the $3+1$ version of (5.2) to be  
$$\eqalign{ S=\int dt\,\int d^3x\,\Bigl\{ &-2\dot{A} _{ai}
\,e M_{ij} e_j{}^a -2 a_i\,D_a\bigl( e M_{ij} e_j{}^a\bigr)\cr 
&+f(\xi ^0)\dot{\xi ^0}\,\eta ^{abc} e_{ai} F_{bci}
-\hat{u} _i\,\eta ^{abc}\bigl[ f(\xi ^0)\partial _a\xi ^0
F_{bci} +\epsilon _{ijk} e_{aj} F_{bck}\bigr]\Bigr\}\,.\cr }\eqno(\z)$$
By using the notation 
$$u^a :=\,{{M_{ij}\hat{u}_j}\over{2e(1+v^2)}} E_i{}^a\,,\eqno(\z)$$
we get for the broken action the expression
$$S=\int dt\,\int d^3x\,\Bigl\{ -\dot{A} _{ai} E_i{}^a  
-a_i D_a E_i{}^a  + u^a\bigl[ C_a -{1\over{4e}} v_a S_{_0}\bigr] 
+2eEf(\xi ^0)\dot{\xi ^0}\bigl[ S_{_0} +2\tilde{v}^a C_a\bigr]
\Bigr\} \,,\eqno(\z)$$
being $\tilde{v}^a$ given by (4.22), and $v_a$ a shorthand for (5.1).
Finally we construct the corresponding Hamiltonian 
$$\eqalign{ H=\int d^3x\,\Bigl\{ &a_i D_a E_i{}^a  -u^a \bigl[
C_a -{1\over{4e}} v_a S_{_0}\bigr] +\lambda _{ai}\bigl( \pi
_i{}^a +E_i{}^a\bigr)\cr 
&+\lambda _i \pi _i  +\mu _i{}^a \sigma _{ia} +\mu ^a \sigma _a 
+\alpha\bigl[ p -2eEf(\xi ^0)\bigl( S_{_0} +2\tilde{v}^a
C_a\bigr) \bigr]\Bigr\}\,,\cr }\eqno(\z)$$
with the momenta defined as in (4.11) and (4.23) respectively. The
essential result, to be directly read out from (5.6), is the set of
first class constraints 
$$\eqalign {&D_a E_i{}^a \approx 0\cr 
&C_a -{1\over{4e}} v_a S_{_0} \approx 0\cr 
&p -2eEf(\xi ^0)\bigl( S_{_0} +2\tilde{v}^a C_a\bigr)\approx 0\,,\cr }
\eqno(\z)$$
with the time component of the tetrad being given by (5.1). Other
constraints are present, identical with those calculated
previously, namely 
$$\pi _i{}^a +E_i{}^a \approx 0\,,\quad \pi _i \approx 0,\quad
\sigma _{ia} \approx 0\,,\quad \sigma _a \approx 0\,.\eqno(\z)$$
Notice that neither $\lambda _i$ nor $\mu ^a\,,\alpha$ can be solved from the stability conditions, thus being associated to first class constraints. We are not interested in these details now. Instead, the most obvious and important consideration to be remarked in a theory with broken time-translational invariance, is that the time operator $p\neq 0$ is no more the generator of a symmetry. Notice that, just as a consequence of the breaking
of the translational symmetry, a Schr\"odinger-type constraint
(5.7c) is present. Contrarily to (4.24b), in principle it is nontrivial,
since neither $C_a$ nor $S_{_0}$ vanish. Indeed, from (5.7b) follows  
$$2\tilde{v}^a C_a \approx v^2 S_{_0}\,.\eqno(\z)$$
When substituted into the Schr\"odinger-like equation (5.7c), the
latter reduces to 
$$\tilde{S}_{_0}:=p -{{f(\xi ^0)}\over{4e}} S_{_0}\approx 0\,.\eqno(\z)$$
Here, the nonvanishing momentum $p$ is a time operator, proportional to the scalar constraint. Eq.(5.10) may be understood as defining a modified scalar
constraint $\tilde{S}_{_0}$. Although now involving $p$, it still retains its original meaning as the generator of reparametrization invariance, a symmetry which remains unbroken. Let us see in particular how it affects the time {\it Goldstone-coordinate} $\xi ^0$. We find $\delta\xi ^0
=\epsilon ^0\left\{ \xi ^0\,,\tilde{S}_{_0}\right\} =\epsilon ^0$, with
$\epsilon ^0$ an arbitrary parameter, leading from $\xi ^0$ to
$\xi ^0 +\delta\xi ^0 =\xi ^0 -\epsilon ^0$. Taking $\epsilon ^0=\xi ^0$, the time variable can be cancelled out. By fixing this gauge, the dynamical content of the theory reduces to the constraints $D_a E_i{}^a =0\,,\, C_a=0$, which are precisely those derived in the Husain-Kuchar model. Actually, in Ref.(16) already we showed that, for $f\left(\xi ^0\,\right) =1$, the resulting dynamics is that corresponding to the Husain-Kuchar case. Now we see that this result generalizes to $f\left(\xi ^0\,\right)\neq 1$. Both alternative choices of $f\left(\xi ^0\,\right)$ yield the same Husain-Kuchar theory. That is not surprising, since a single scalar field defines a foliation, being a solution of the Frobenius condition. Accordingly, the Husain-Kuchar model admits an interesting interpretation in terms of the theory we are discussing here, namely as ordinary gravity with broken time-translational symmetry, see Ref.(16). 

In contrast to the translational symmetry, the reparametrization invariance remains a symmetry of the theory. Actually, this symmetry is responsible for the following peculiarity. In principle, our model depends on a time-like variable $\xi ^0$. However, such variable can be gauged away by means of a reparametrization transformation. In this sense, we say
that time evolution defined by (5.2) is gauge dependent. Let us call this fictitious sort of "time evolution" a gauge dependent dynamical evolution. Obviously, since a gauge choice exists for which the theory becomes independent on $\xi ^0$, the current model cannot be a satisfactory characterization of physical time evolution. Thus, the attempts presented by us until now as candidates to characterize nontrivial dynamical time evolution in the context of dynamical spacetimes, still remain unsatisfactory. Perhaps, one could suspect that the difficulty in defining time evolution in the context of the present model could derive from the drastic procedure by which, in it, the translational symmetry is broken, consisting in simply putting the translational connection $\Gamma _a$ in (4.17) equal to zero. Thus, it is reasonable to explore other ways of breaking the symmetry, respecting at the same time the translational connection as a nonvanishing field. As we will see, this does not solve the problem. 

Let us introduce an auxiliary metric $g^{ab}$ allowing to define transversality. For the moment we do not prejudge if it has to be considered as an additional dynamical field, or not. In order to decompose the connection into a longitudinal and a transversal part, we add to the Lagrangian a term of the form $\lambda\,g^{ab}\partial _a\Gamma _b$, with $\lambda$ a Lagrange multiplier. If one takes $g^{ab}$ to be the dynamical metric tensor, constructed from the {\it vierbeine} as $g^{ab}=e^a{}_I e^b{}_I$, after tedious calculations one finds essentially the same result as before. Indeed, a first class scalar constraint is obtained, with the same structure as (5.7c), namely $p$ plus terms not contributing to the Poisson bracket with $\xi ^0$. Exactly as in the case studied above, reparametrizations remain unbroken, and no real time evolution follows, but gauge dependent time evolution. So, nothing new is achieved in this way. On the other hand, being the auxiliary metric mainly a tool to implement the notion of transversality, one could try to break the reparametrization invariance by considering $g^{ab}$ as a non dynamical, or even as a constant object. Contrarily to what one would expect from a theory where both, the time-translational and the reparametrization invariance are broken, calculations show that, once more, an expression consisting of $p$ plus terms not contributing to the Poisson bracket with $\xi ^0$, is obtained. Although no more interpretable as the generator of one of the original symmetries, even in this case a scalar first class constraint, with an arbitrary Lagrange multiplier, is present, reproducing the previous scheme, and thus allowing the time variable to be gauged away. 
\bigskip\bigskip

\sectio{\bf Gauge independent time evolution with gravitation.}\bigskip 

From the previous dicussion, we conclude that a certain feature of the action (4.1) of ordinary gravity seems to constitute an obstacle to break time-translational invariance in such a way that it becomes totally disentangled from any other --essential or accidental-- symmetry of the action. The resulting apparent time evolution remains in fact gauge dependent in all the cases considered by us above. As we have seen, standard gravity with the additional condition of broken time-translational invariance, after eliminating the time variable by means of a gauge fixing, yields the Husain-Kuchar model; and the same result follows from alternative breaking methods based on the obvious decomposition of the time-translational connection into longitudinal and transversal parts. Since all these breaking mechanisms yield analogous results, we suspect that the problem in avoiding the possibility of gauging away the time variable probably derives from the structure of the action of ordinary GR itself. If that were the case, then, for time evolution to be meaningfully defined, the gravitational action should be modified to some extent in order to remove the theoretical difficulties. We propose to consider, as a natural generalization of standard gravity, the Jordan-Brans-Dicke$^{(18)}$ (JBD) action
$$S={1\over{16\pi G}}\int d^4x\sqrt{-g}\left(\varphi R -{\omega\over{\varphi}}\,g^{ab}\partial _a\varphi 
\partial _b\varphi\,\right)\,.\eqno(\z)$$
The occurrence of a scalar field at low energy effective gravity also receives an independent support from string theories. Our interest on the JBD action (6.1) is motivated by the fact that it is a good candidate to constitute a starting point, different from standard gravity, with the formal features we need to break down time-translational symmetry in a more suitable way as what we studied in the previous section. The inclusion of the scalar field contribution, as a new constituent of the gravitational action, opens the question on its interpretation. In particular, we wonder if the scalar field has necessarily to be a new field, rather than one of the old fields already contained in the action of ordinary gravity. Indeed, our proposal consists in identifying the JBD scalar field $\varphi$ with the time {\it Goldstone-coordinate} $\xi ^0$ introduced by us, up to dimensional factors that can be absorbed in a redefined $\omega$ in (6.1). From this point of view, (6.1) becomes interpretable as an action with broken time-translational invariance, whereas the remaining symmetries are respected. This is a consequence of the fact that the time variable $\xi ^0$ can be chosen to be a nonlinear $SO(4)$ or Lorentz invariant, however transforming under time translations. (The possibility of modifying the transformation properties of $\xi ^0$ rests on the flexibility in passing from a NLR to another with a different structure subgroup $H\subset G$. See in particular Ref.(8).) 

The Hamiltonian treatment of the JBD action has been studied by Garay and Garc\'ia-Bellido$^{(19)}$. Starting from the action (6.1) rewritten in the Einstein frame, and following the standard ADM description, a singular Hamiltonian is obtained consisting of a linear combination of constraints where both, the vector and scalar constraints, compare with (4.15, 4.16), are present, namely 
$$C_a =-2\,q_{ab}D_c\,p^{bc} +p\,\partial _a\varphi\,,\eqno(\z)$$
and 
$$S_{_0} ={{16\pi G}\over{\sqrt{q}}}\left[\, p^{ab}p_{ab} -{1\over2}\left( q^{ab}p_{ab}\,\right) ^2 +{1\over2}\,p^2\,\right] +{{\sqrt{q}}\over{16\pi G}}\left( R^{(3)} -q^{ab}\partial _a\varphi\partial _b\varphi\,\right)\,.\eqno(\z)$$
In (6.2) and (6.3), the time momentum conjugate to the scalar field $\varphi $ identified by us with the time variable $\xi ^0$, is denoted by $p$, as before, whereas $q^{ab}$ is the three-metric, with $p_{ab}$ as its conjugate momentum, and $R^{(3)}$ stands for the three-dimensional curvature. From the new point of view proposed by us, the interesting thing is the occurrence in (6.3) of the quadratic term $p^2$ in time momentum, in analogy to the free particle action. In virtue of the presence of such quadratic term in the scalar constraint, the time variable $\xi ^0$ cannot be gauged away anymore. This essentially modifies the ordinary theory of gravity, in the sense that a non pathological time evolution becomes allowed to coexist with an observationally well tested dynamical theory of spacetime. Actually, phaenomenological estimations of the JBD parameter $\omega$ in (6.1) exist, turning out to be very large, and thus far from the critical conformal value in which the kinetic term for $\varphi$ could be reabsorbed into the remaining fields by means of a reparametrization transformation. Thus, the observational arguments favouring the unavoidability of this kinetic term in the action, automatically support the existence of the quadratic term $p^2$ in (6.3),  derived from that contribution.  
\bigskip\bigskip 

\sectio{\bf Final remarks.}
\bigskip 

By applying the three postulates established by us in the present paper to the NLR-approach to gauge theories of gravity, a certain conception of time evolution follows in the presence of gravitation. Let us shortly review how this is achieved. First of all, we consider a gravitational gauge theory {\`a la Hehl}, for instance PGT, where the translations are considered as a constitutive part of the internal gauge group. Important is, with regard to that group, to make use of the nonlinear machinery, in order to get for the tetrad the unitary-gauge structure (2.9), where the {\it Goldstone coordinates} occur. 

Then, the breaking of the translational time symmetry, as required by the first postulate, provides us with a quite natural internal time variable, namely the time {\it Goldstone coordinate} $\xi ^0$, whose changes become physical as a consequence of the symmetry breaking. On the other hand, according to the second postulate, physical laws express correlations between dynamical variables. In particular, time evolution is then to be defined with respect to the already identified internal time $\xi ^0$. Of course, there is no problem in expressing these correlations parametrically, although it can be somewhat misleading, since evolution could be confused with reparametrizations with respect to the non-physical affine parameter. In any case, physically meaningful information can actually be read out from the Poisson brackets of any dynamical variable with the singular Hamiltonian. This is possible due to the fact that physical processes involve {\it reparametrization invariant time evolution}$^{(3)}$, evaluated with respect to the time {\it Goldstone-coordinate} $\xi ^0$ introduced previously. 

At this point, we tried to get nontrivial time evolution from an action of standard gravity. Nevertheless, although we broke down the translational symmetry in different forms, the interplay between the time-translational symmetry and the scale invariance symmetry always yielded a surviving scalar constraint allowing to gauge away the time variable from the theory. In these cases, the resulting action revealed to be equivalent to the Husain-Kuchar model, the latter thus being interpretable as ordinary gravity with broken time-translational symmetry. Finally, our various unssuccessful attempts to get time evolution with respect to internal time from ordinary gravity lead us to disregard it in favour of a JBD action with the scalar field $\varphi$ identified with the time field $\xi ^0$. 

Let us next examine that action in light of the third postulate, which requires the special relativistic limit to be recovered when gravity is switched off. Certainly, it is not evident that the scalar field kinetic term trivializes in the absence of gravity. That constitutes a question to be interpreted in the JBD description, as far as the scalar field $\varphi$ is considered as an additional degree of freedom which is not present in standard gravity, its dynamics persisting in principle in a Minkowskian spacetime. Nevertheless, without necessity of requiring any additional condition on $\varphi$ in the Minkowskian limit, the problem is automatically solved by our assumption that $\varphi$ is to be identified with the time variable $\xi ^0$, as will become clear from the following observation on the meaning of the JBD action (6.1) in the limit of vanishing gravitational interaction. 

One expects that, by switching off gravity, the usual dynamics on Minkowski spacetime should be recovered. Notice that in the limit (3.2) of absence of gravity, the time-translational, as much as the space-translational and the reamaining spacetime connections, vanish, so that $v_a =\partial _a\xi ^0$ and $e_{ai} =\partial _a\xi ^i$. With our identification $\varphi =\xi ^0$, simple calculations show the kinetic term of $\varphi$ of the action (6.1) in the Einstein frame, namely $d^4 x\,\sqrt{-g}\,g^{ab}\partial _a\varphi \partial _b\varphi = d^4 x\,e\,g^{ab}\partial _a\xi ^0 \partial _b\xi ^0 $, to reduce to $d^4 x\,det\left( \partial _a\xi ^\alpha\,\right)$ (where $\alpha$ runs over $0$ and $i$) and thus, trivially, to $d^4 \xi$. See also Ref.(16). So, in the limit of vanishing gravity, the JBD term becomes a constant, not contributing to the dynamics. (Observe the coincidence with the corresponding limit of the cosmological constant term.) The special relativistic limit is thus reached. 

The latter feature does not automatically follow from the standard JBD action; rather it is a consequence of the interpretation of the scalar field proposed by us as $\varphi =\xi ^0$. Indeed, this additional hypothesis concerning the particular JBD model --widely accepted as a natural extension of the Einstein-Hilbert action-- guarantees that our three postulates on time evolution are satisfied by it. However, we do not advocate for the JBD action, even enlarged with additional potentials, as the only possible candidate to replace GR. We only claim that it possesses several desirable features required by us to be satisfied by any satisfactory gravitational action. The search for such an action should throw some new light on the general problem of time evolution in the presence of gravitational interaction. 
\bigskip\bigskip 

\centerline{\bf Acknowledgement}
\bigskip 
We are very grateful to Dr. J.F. Barbero for helpful discussions. 
\bigskip

\bigskip\bigskip
\centerline{REFERENCES}
\bigskip 
\noindent\item{(1)} C.J. Isham, {\it Canonical Quantum Gravity
and the Problem of Time}, in Recent Problems in Mathematical
Physics, Salamanca, June 15--27, 1992

\item\qquad C. Kiefer, Report Freiburg THEP-94/4, {\it Contribution to the Lanczos Conference Proceedings}, (gr-qc/9405039)

\item\qquad J.B. Hartle, {\it Class. Quantum Grav.} {\bf 13} (1996) 361

\noindent\item{(2)} K.V. Kuchar, in {\it Proceedings of the 4th
Canadian Conference on General Relativity and Relativistic
Asrtophysics}, eds. G. Kunstatter, D. Vincent and J. Williams
(World Scientific, Singapore, 1992), and references therein

\noindent\item{(3)} C. Teitelboim, F. Ruiz Ruiz and A. Gonz\'alez L\'opez, {\it Anales F\'is.} {\bf A 80} (1984) 58  

\item\qquad A. Hanson, T. Regge and C. Teitelboim, {\it
Constrained Hamiltonian Systems}, Roma, Accademia dei Lincei
(1976)

\noindent\item{(4)} C. Rovelli, {\it Phys. Rev.} {\bf D 42}
(1990) 2638, {\it Nuov. Cim.} {\bf 110} (1995) 81

\noindent\item{(5)} A. Anderson, gr-qc/9507039

\noindent\item{(6)} C. Rovelli, {\it Phys. Rev.} {\bf D 43}
(1991) 442

\noindent\item{(7)} J. Julve, A. L\'opez--Pinto, A. Tiemblo and R.
Tresguerres, {\it G.R.G.} {\bf 28} (1996) 759

\item\qquad A. L\'opez--Pinto, A. Tiemblo and R. Tresguerres, 
{\it Class. Quantum Grav.} {\bf 12} (1995) 1503

\item\qquad A. L\'opez--Pinto, A. Tiemblo and R. Tresguerres, 
{\it Class. Quantum Grav.} {\bf 13} (1996) 2255

\item\qquad R. Tresguerres and E.W. Mielke, {\it Phys. Rev.} {\bf D 62} (2000) 044004

\noindent\item{(8)} A. L\'opez--Pinto, A. Tiemblo and R. Tresguerres, 
{\it{Hamiltonian Poincar\'e Gauge Theory of Gravitation}}, 
{\it Class. Quantum Grav.} {\bf 14} (1997) 549

\noindent\item{(9)} A. Tiemblo and R. Tresguerres, 
{\it{Invariant foliation of dynamical spacetimes}}

\noindent\item{(10)} S. Coleman, J. Wess and B. Zumino, {\it Phys. Rev.}
{\bf 117} (1969) 2239 

\item\qquad C.G. Callan, S. Coleman, J. Wess and B. Zumino, {\it Phys. Rev.}
{\bf 117} (1969) 2247

\item\qquad S. Coleman, {\it Aspects of Symmetry}. Cambridge
University Press, Cambridge (1985)

\item\qquad A.B. Borisov and I.V. Polubarinov, {\it Zh.
ksp. Theor. Fiz.} {\bf 48} (1965) 1625, and V. Ogievetsky and 
I. Polubarinov, {\it Ann. Phys.} (NY) {\bf 35} (1965) 167

\item\qquad A.B. Borisov and V.I. Ogievetskii, {\it Theor.
Mat. Fiz.} {\bf 21} (1974) 329

\item\qquad A. Salam and J. Strathdee, {\it Phys. Rev.}
{\bf 184} (1969) 1750 and 1760

\item\qquad C.J. Isham, A. Salam and J. Strathdee, {\it Ann. of Phys.} 
{\bf 62} (1971) 98

\item\qquad L.N. Chang and F. Mansouri, {\it Phys.
Lett.} {\bf 78 B} (1979) 274, and {\it Phys. Rev.} {\bf D 17} (1978) 3168

\item\qquad K.S. Stelle and P.C. West, {\it Phys. Rev.} 
{\bf D 21} (1980) 1466

\item\qquad A.A. Tseytlin, {\it Phys. Rev.} {\bf D 26} (1982) 3327

\item\qquad E.A. Lord, {\it{Gen. Rel. Grav.}} {\bf 19} (1987) 983, and 
{\it J. Math. Phys.} {\bf 29} (1988) 258

\noindent\item{(11)} R. Utiyama,  {\it Phys. Rev.} {\bf 101} (1956) 1597

\item\qquad T. W. B. Kibble, {\it J. Math. Phys.} {\bf 2} (1961) 212

\item\qquad D. W. Sciama, {\it Rev. Mod. Phys.} {\bf 36} (1964) 463 and 1103

\item\qquad K. Hayashi and T. Nakano, {\it Prog. Theor. Phys} 
{\bf 38} (1967) 491

\item\qquad A. Trautman, in {\it Differential Geometry},
Symposia Mathematica Vol. 12 (Academic Press, London, 1973), p. 139

\item\qquad A. G. Agnese and P. Calvini, {\it Phys. Rev.} {\bf
D 12} (1975) 3800 and 3804 

\item\qquad E.A. Ivanov and J. Niederle, {\it Phys. Rev.} {\bf D25} 
(1982) 976 and 988

\item\qquad D. Ivanenko and G.A. Sardanashvily, {\it
Phys. Rep.} {\bf 94} (1983) 1

\item\qquad E. A. Lord, {\it J. Math. Phys.} {\bf 27} (1986) 2415 and 3051

\item\qquad K. Hayashi and T. Shirafuji, {\it Prog. Theor. Phys} 
{\bf 64} (1980) 866 and {\bf 80} (1988) 711

\item\qquad G. Grignani and G. Nardelli, {\it Phys. Rev.} 
{\bf D 45} (1992) 2719

\noindent\item{(12)} F.W. Hehl, P. von der Heyde, G. D. Kerlick
and J. M. Nester, {\it Rev. Mod. Phys.} {\bf 48} (1976) 393

\item\qquad P. von der Heyde, {\it Phys. Lett.} {\bf 58 A} 
(1976) 141

\item\qquad F.W. Hehl, {\it Proc. of the 6th Course of the School of
Cosmology and Gravitation on Spin, Torsion, Rotation, and
Supergravity}, held at Erice, Italy, May 1979, P.G. Bergmann, V.
de Sabbata, eds. (Plenum, N.Y. 1980) 5

\item\qquad F. W. Hehl, J. D. McCrea,  E. W. Mielke, and
Y. Ne'eman {\it Found. Phys.} {\bf 19} (1989) 1075

\item\qquad F.W. Hehl, J.D. McCrea, E.W. Mielke, and Y. Ne'eman,
{\it Physics Reports} {\bf 258} (1995) 1, and references therein

\noindent\item{(13)} A. Ashtekar, {\it Phys. Rev. Lett.} {\bf 57}
(1986) 2244

\item\qquad A. Ashtekar, {\it Phys. Rev.} {\bf D 36} (1987) 1587

\item\qquad A. Ashtekar, {\it Non Perturbative Canonical Gravity}  
(Notes prepared in collaboration with R. S. Tate). (World 
Scientific Books, Singapore, 1991)

\item\qquad J. F. Barbero G., {\it Phys. Rev.} {\bf D 51} (1995)
5507

\noindent\item{(14)} E. W. Mielke, J.D. McCrea, Y. Ne'eman and F.W. Hehl {\it Phys. Rev.} {\bf D 48} (1993) 673

\noindent\item{(15)} C. Rovelli, {\it Nuov. Cim.} {\bf 110 B} (1995) 81

\noindent\item{(16)} V. Husain and K. Kuchar, {\it Phys. Rev.} {\bf D 42} (1990) 4070

\item\qquad J.F. Barbero, A. Tiemblo and R. Tresguerres, {\it Phys. Rev.} {\bf D 57} (1998) 6104

\noindent\item{(17)} J. Samuel, {\it Pramana J. Phys} {\bf 28} (1987) L429 

\item\qquad T. Jacobson and L. Smolin, {\it Phys. Lett.} {\bf B 196} (1987) 39; {\it Class. Quantum Grav.} {\bf 5} (1988) 583

\noindent\item{(18)} P. Jordan, {\it Nature} {\bf 164} (1949) 

\item\qquad P. Jordan, {\it Z. Phys.} {\bf 157} (1959) 112

\item\qquad C. Brans and R.H. Dicke, {\it Phys. Rev.} {\bf 124} (1961) 925

\item\qquad R.H. Dicke, {\it Phys. Rev.} {\bf 125} (1962) 2163

\item\qquad C. Brans, {\it Phys. Rev.} {\bf 125} (1962) 2194

\noindent\item{(19)} L.J. Garay and J. Garc\'ia-Bellido, {\it Nucl. Phys.} {\bf B 400} (1993) 416

\bye